\documentclass[aps,prl,groupedaddress,twocolumn,showpacs]{revtex4-1}
\usepackage{graphicx,psfrag,amsmath,calc,amssymb, comment, color}

\begin{document}

\title{Density induced BCS-Bose evolution in gated two-dimensional superconductors: \\
The Berezinskii-Kosterlitz-Thouless transition as a function of carrier density}

\author{Tingting Shi$^{1,2}$}
\author{Wei Zhang$^{1}$}
\thanks{wzhangl@ruc.edu.cn}
\author{C. A. R. S{\' a} de Melo$^{2}$}
\thanks{carlos.sademelo@physics.gatech.edu}
\affiliation{$^{1}$ Department of Physics, Renmin University of China, Beijing 100872, China}
\affiliation{$^{2}$ School of Physics, Georgia Institute of Technology,
Atlanta, Georgia 30332, USA}


\begin{abstract}
We discuss the evolution from BCS to Bose superconductivity versus carrier density in
gated two-dimensional $s$-wave superconductors. We investigate the carrier density
dependence of the critical temperature, superfluid density, order parameter, chemical
potential and pair size. We show that
the transition from the normal to the superconducting state is controlled by the
Berezinskii-Kosterlitz-Thouless vortex-antivortex pairing mechanism, and that the evolution
from high carrier density (BCS pairing) to low carrier density (Bose pairing)
is just a crossover in $s$-wave systems. We compare our results to 
recent experiments on the superconductor ${\rm Li}_{x}{\rm ZrNCl}$,
a lithium-intercalated layered nitride, and obtain very good quantitative agreement
at low and intermediate densities.
\end{abstract}

\maketitle

{\it Introduction:}
The physics of two-dimensional (2D) superconductors is a fascinating subject with
a long history~\cite{goldman-2013}.
The theoretical prediction that 2D Fermi systems can exhibit superconductivity
can be inferred from the seminal papers of Berezinskii~\cite{berezinskii-1970}, 
Kosterlitz and Thouless~\cite{kosterlitz-thouless-1972}, where the mechanism of
vortex-antivortex unbinding was proposed.
The topological nature of the superconducting transition in two dimensions
at fixed carrier density was unveiled in experimental studies of thin films
~\cite{beasley-1979, goldman-1981}, which provided strong evidence for the
Berezinskii-Kosterlitz-Thouless (BKT) transition and
vortex-antivortex unbinding~\cite{halperin-1979}.

Chemical substitutions or atom deficiencies in 2D materials
can change carrier density, but very often they also introduce disorder
and inhomogenous doping, creating difficulties in the interpretration of experimental
data and phase diagrams. A way to circumvent such difficulties is to study 2D
systems via electrostatic doping, where the carrier density $n$ can be tuned from
a minimum $n_{\rm min}$ to a maximum $n_{\rm max}$ value. It was theoretically
proposed that topological quantum phase transitions occur during the
evolution from BCS to Bose pairing in 2D $d$- and $p$-wave
superfluids and superconductors as a function of density or interaction
strength~\cite{duncan-2000, botelho-2005a, read-2000, botelho-2005b}.
However, experiments that used electrostatic doping in 2D
$d$-wave superconductors (cuprates)~\cite{ahn-2003a, ahn-2003b}
did not investigate the existence of a quantum critical point from gapless to gapped
from high (BCS pairing) to low (Bose pairing) density of carriers,
but indicated that phase fluctuations play an important role
in determining the critical temperature at low carrier densities.
This experimental observation was in
agreement with a theoretical argument that classical phase fluctuations
were very important in establishing an upper bound for the critical temperature of
2D superconductors with small superfluid density~\cite{kivelson-1995}.
Furthermore, these early experiments provided support to the viewpoint
that the relation between the critical temperature $T_c$ and the muon depolarization rate
in cuprate superconductors~\cite{uemura-1989, uemura-1991}
should be reinterpreted as an upper bound on $T_c$ given by the ordering
temperature for phase fluctuations~\cite{kivelson-1995}.

For single band 2D superfluids and superconductors with parabolic bands,
it was shown explicitly that this upper bound for the critical temperature
$T_{c} = T_{\rm BKT}$ is exactly one-eighth of the Fermi energy $\varepsilon_F$,
that is, $T_{\rm BKT} \le \varepsilon_F/8 = 0.125 \varepsilon_F$~\cite{botelho-2006}
in units where $\hbar = k_B = 1$. Such a general relation is obtained rather
trivially from the BKT transition temperature $T_{\rm BKT} = \pi \rho_s (T_{\rm BKT}) /2$,
where $\rho_s$ is the superfluid density. It is sufficient to notice that, at any temperature
$T$, $\rho_s (T)$ cannot exceed $n/(4m)$, where $n$ is the carrier density and $m$
is the carrier effective mass. Physically, this bound is the ratio between the maximum
fermion pair density $(n/2)$ and the fermion pair mass $2m$ and is reflects
the Ferrell-Glover-Tinkham~\cite{ferrell-1958, tinkham-1959, tinkham-1975} optical conductivity sum rule.
The same bound is valid even when the superfluid density tensor
is anisotropic or when spin-orbit effects are included~\cite{devreese-2014, devreese-2015}.

Experiments seeking to study the evolution from BCS to Bose pairing in essentially
2D superconductors like FeSe~\cite{kasahara-2014, rinott-2017, hashimoto-2020},
twisted bilayer graphene~\cite{cao-2018, herrero-2021}, and 
layered nitrides~\cite{nakagawa-2018} have been attempted, but only very 
recently it has been shown that in ${\rm Li}_{x}{\rm ZrNCl}$, a lithium-intercalated layered nitride,
the electrolyte ${\rm LiClO_{4}}$ can dope a 2D  layer
of ${\rm ZrNCl}$ with Lithium ions via a gate voltage $V_G$,
over a range of nearly two orders of magnitude~\cite{iwasa-2021}.
Because of this doping tunability, this technique
potentially alllows for the study of the evolution from BCS (high carrier density)
to the Bose (low carrier density) superconductivity in 2D systems
as discussed theoretically in the context of ultracold fermions~\cite{botelho-2006}.
In this paper, we show that a theory involving
phase fluctuations for 2D $s$-wave superconductors
leads to critical temperature and order parameter modulus
in good agreement with experimental results
for ${\rm Li}_{x}{\rm ZrNCl}$~\cite{iwasa-2021} as the system evolves from BCS to Bose
pairing in the relevant carrier concentration range.
Furthermore, we extract from experimental data the effective mass of the charge carriers,
as well as, the magnitude and range of the effective two-body interaction between them.

{\it Hamiltonian:}
We use a two-band continuum model of fermion
with identical parabolic bands centered at points $K$ and $K^\prime$ of the Brillouin zone of
${\rm Li}_{x}{\rm ZrNCl}$. Since the two bands are identical, the effective mass of the carriers
in each band is $m$ and density of carriers per band is $n = k_F^2 / 2\pi$.
Furthermore, we take the interactions to be the same in both bands
and assume that the two bands are independent. In this case, it is sufficient to consider
the Hamiltonian per band. Thus, we start from the Hamiltonian density per band ($\hbar=k_B=1$) 
$
{\cal H} ({\bf r})
=
{\cal H}_{\rm K} ({\bf r})
+
{\cal H}_{\rm I} ({\bf r}), 
$
where the kinetic energy density with respect to the chemical
potential $\mu$ is
\begin{equation}
\label{eqn:hamitonian-kinetic-energy}
{\cal H}_{\rm K} ({\bf r})
=
\sum_s
\psi^\dagger_{s} ({\bf r}) \left[  -\frac{\nabla^2}{2m}  - \mu \right] \psi_s ({\bf r}),
\end{equation}
and the interaction energy density is
\begin{equation}
\label{eqn:hamiltonian-interaction}
{\cal H}_{\rm I} ({\bf r})
=
\int d^2 {\bf r}^\prime
V ({\bf r}, {\bf r}^\prime)
\psi_{\uparrow}^\dagger ({\bf r}) \psi_{\downarrow}^\dagger ({\bf r}^\prime)
\psi_{\downarrow} ({\bf r}^\prime)  \psi_{\uparrow} ({\bf r})
\end{equation}
with
$V ({\bf r}, {\bf r}^\prime) = - V_s g (\vert {\bf r} - {\bf r}^\prime\vert/ R)$.
Here, $V_s$ is the magnitude of the $s$-wave attractive interaction with dimension
of energy, and $g (\vert {\bf r} - {\bf r}^\prime\vert/ R)$ is a dimensionless function
with range $R$. The field operator $\psi_s^{\dagger} ({\bf r})$ creates a fermion
with spin projection $s$ at position ${\bf r}$.
The Hamiltonian $H = \int d^2 {\bf r} {\cal H} ({\bf r})$ has three microscopic
parameters: the effective mass $m$, the interaction strength $V_s$ and
the interaction range $R$. In momentum space, the Hamiltonian reads
\begin{equation}
\label{eqn:hamiltonian-momentum-space}
H =
 \sum_{{\bf k}, s} 
\xi_{\bf k} \psi_{{\bf k}, s}^\dagger \psi_{{\bf k}, s} + 
\sum_{{\bf k},{\bf k}',{\bf q}} V_{{\bf k}{\bf k}'}
b_{{\bf k}{\bf q}}^\dagger b_{{\bf k}'{\bf q}} ,
\end{equation}
where 
$
b_{{\bf k}{\bf q}} = \psi_{-{\bf k}+{\bf q}/2, \downarrow}
\psi_{{\bf k}+{\bf q}/2, \uparrow}$ is the pairing operator 
and
$\xi_{\bf k}= \varepsilon_{\bf k}  - \mu$, 
with energy dispersion $\varepsilon_{\bf k} = {\bf k}^2/2m$.
Following standard procedure~\cite{duncan-2000}, we obtain a separable
interparticle potential in momentum space
\begin{equation}
\label{potential}
V_{{\bf k}{\bf k}'} = -V_s 
\Gamma_s ({\bf k}) \Gamma_s({\bf k}') ,
\end{equation}
where
$
\Gamma_s ({\bf k}) = 
\left( 1 + k/k_R \right)^{-1/2},
$
and $k_R \sim  R^{-1} $ plays the role of the interaction range in momentum space~\cite{footnote-1}.
The Hamiltonian in Eq.~(\ref{eqn:hamiltonian-momentum-space}) has three microsopic parameters:
the effective mass $m$, the interaction strength $V_s$ and range
$R \sim k_R^{-1}$.
As the density $n$ is varied, we compare the interparticle spacing
$\ell_{\rm ip} = 1/\sqrt{n}= \sqrt{2\pi}/k_F$ to $R$ and notice that the interactions are
short-ranged when 
$R/\ell_{\rm ip} \sim (k_F/k_R)(1/\sqrt{2\pi}) \ll 1$
and long-ranged when
$R/\ell_{\rm ip} \sim (k_F/k_R)(1/\sqrt{2\pi}) \gg 1$.

{\it Effective Action:}
We consider the order parameter modulus $\vert \Delta \vert$ and its associated phase $\theta$,
to write the phase fluctuation effective action per band as~\cite{botelho-2006}
\begin{equation}
S_{\rm eff} =
S_{\rm sp} \left( \vert \Delta \vert \right)
+ 
S_{\rm ph} ( \vert \Delta \vert, \theta)
+
S_{\rm zp} ({\vert \Delta \vert, \theta})
\end{equation}
The first term is the saddle-point action 
\begin{equation}
\label{eqn:saddle-point-action}
S_{\rm sp} \left( \vert \Delta \vert \right)
=
\sum_{\bf k}
\left[
\frac{\left( \xi_{\bf k} - E_{\bf k} \right)}{T}
-
2 \ln \left(  1 + e^{-E_{\bf k}/T}\right)
\right] + \frac{\vert \Delta \vert^2}{T V_s},
\end{equation}
where
$
E_{\bf k} = \sqrt{\xi_{\bf k}^2 + |\Delta_{\bf k}|^2}
$
is the quasiparticle excitation energy, and 
$\Delta_{\bf k} = \Delta \Gamma_{s} ({\bf k})$ plays the role
of the order parameter function for $s$-wave pairing.
The second term is the phase fluctuation action 
\begin{equation}
\label{eqn:phase-fluctuation-action}
S_{\rm ph} = \frac {1} {2} \int dr \,
\left[
\sum_{ij} \rho_{ij} \, \partial_i \theta(r) \partial_j \theta(r)
+
\kappa_s \left[ \partial_{\tau} \theta (r) \right]^2 
\right],
\end{equation}
where the integration is over position and imaginary time $r = ({\bf r}, \tau)$, with
$\int dr \equiv \int_0^{1/T} d\tau \int d^2{\bf r}$.
In Eq.~(\ref{eqn:phase-fluctuation-action}), the first term contains the superfluid density tensor
\begin{equation}
\label{eqn:superfluid-density}
\rho_{ij}(\mu, \vert \Delta \vert, T)
=
\frac {1} {4 L^2} \sum_{\bf k}
\left[
2 n_{\rm sp} ({\bf k}) \partial_i \partial_j \xi_{\bf k} - 
Y_{\bf k} \partial_i \xi_{\bf k} \partial_j \xi_{\bf k}
\right] ,
\end{equation}
where $\partial_i$ denotes the partial 
derivative with respect to momentum $k_i$ with $i = \{ x, y \}$,
\begin{equation}
\label{eqn:momentum-distribution}
n_{\rm sp} ({\bf k}) = \frac {1} {2} \left[
1 - \frac {\xi_{\bf k}} {E_{\bf k}} 
\tanh \left( \frac { E_{\bf k} } {2 T} \right) \right]
\end{equation}
is the momentum distribution per spin state,
and
$
Y_{\bf k} = (2T)^{-1} 
{\rm sech}^2 (E_{\bf k} / 2T)
$
is the Yoshida function.
Here, the superfluid density tensor is diagonal
since $\rho_{xy} = \rho_{yx} = 0$, and its diagonal elements
are equal, that is, $\rho_{xx} = \rho_{yy} = \rho_s$.
The second term in Eq.~(\ref{eqn:phase-fluctuation-action}) is 
\begin{equation}
\label{eqn:compressibiity}
\kappa_{s} (\mu, \vert \Delta \vert, T)
=
\frac {1} {4 L^2} \sum_{\bf k} 
\left[
\frac {|\Delta_{\bf k}|^2} { E_{\bf k}^3 }
\tanh \left( \frac { E_{\bf k} }  { 2 T}  \right) + 
\frac { \xi_{\bf k}^2 } { E_{\bf k}^2 } Y_{\bf k}
\right] ,
\end{equation}
with $\kappa_s = \kappa/4$, where
$\kappa = \partial n/ \partial \mu \vert_{T,V}$ is related to the
thermodynamic compressibility ${\cal K} = \kappa/n^2$.
The  zero-point action is $S_{\rm zp} = \sum_{{\bf q}} \omega_{\bf q}/2T$,
where $\omega ({\bf q}) = c |{\bf q}|$ is the frequency and 
$c = \sqrt {\rho_s/\kappa_{s}}$ is the speed of sound. 

To elucidate the role of phase fluctuations, we write
$\theta ({\bf r}, \tau) = \theta_v ({\bf r}) + \theta_c ({\bf r}, \tau)$, 
where the classical ($\tau$-independent) contribution $\theta_v ({\bf r})$ is due to vortices 
(transverse velocities), while the quantum ($\tau$-dependent) contribution
$\theta_c ({\bf r}, \tau)$ is due to collective modes (longitudinal velocities).
The resulting phase action per band is
\begin{equation}
\label{eqn:vortex-collective-action}
S_{\rm ph} = S_{v} + S_{c},
\end{equation}
since contributions involving cross-terms in $\theta_v$ and $\theta_c$ vanish.
The vortex action is  
\begin{equation}
\label{eqn:vortex-action}
S_{v}
=
\frac{1}{2T} \int d^2{\bf r} \,
\rho_s\left[ \nabla \theta_v ({\bf r}) \right]^2,
\end{equation}
while the collective mode action is 
\begin{equation}
\label{eqn:collective-action}
S_{c} = \frac{1}{2} \int dr 
\left[
\rho_s \left[ \nabla \theta_{c} (r) \right]^2
+
\kappa_{s} \left[ \partial_\tau \theta_{c} (r) \right]^2
\right].
\end{equation}

The thermodynamic potential is
$\Omega= \Omega_{\rm sp} + \Omega_{\rm ph}$,
where the first term is the saddle-point
$\Omega_{\rm sp}= T S_{\rm sp}$,
and the second is due to phase fluctuations
$\Omega_{\rm ph} = \Omega_{v} + \Omega_c + \Omega_{\rm zp}$. 
The contribution due to vortices is $\Omega_{v} = -T \ln {\cal Z}_v$, where
${\cal Z}_v = \int d\theta_v e^{-S_v}$.
Integration over phase fluctuations $\theta_c ({\bf r}, \tau)$ leads to 
$
\Omega_{c} = \sum_{\bf q} 
T\ln \left[ 1 - \exp \left( -  { \omega_{\bf q} / T } \right) \right],
$
while the zero-point term is $\Omega_{\rm zp} = T S_{\rm zp}$.

{\it Critical Temperature:} The self-consistency relations for $\vert \Delta \vert$
and $T_c$ for fixed chemical potential $\mu$ are derived from the
effective action $S_{\rm eff} = S_{\rm sp} + S_{\rm ph}$ as follows.
The order parameter equation is obtained through
the stationarity condition 
$
\delta S_{\rm sp} / 
\delta \Delta^*_0 = 0 ,
$
leading to 
\begin{equation}
\label{eqn:order-parameter}
\frac {1} {V_{s}}  = 
\sum_{\bf k} \frac {|\Gamma_{s} ({\bf k})|^2} {2 E_{\bf k}}
\tanh \left( \frac {E_{\bf k}} {2 T} \right) .
\end{equation}

The equation for the critical temperature $T_c = T_{\rm BKT}$ 
is determined by the Kosterlitz-Thouless~\cite{kosterlitz-thouless-1972} condition
\begin{equation}
\label{eqn:critical-temperature}
T_{\rm BKT} = \frac {\pi} {2} \rho_s ( \mu, \vert \Delta \vert, T_{\rm BKT} ),
\end{equation}
obtained by using the vortex action of Eq.~(\ref{eqn:vortex-action}).
For fixed interaction $V_s$ and a given chemical potential $\mu$,
Eqs.~(\ref{eqn:order-parameter}) and~(\ref{eqn:critical-temperature})
determine the order parameter modulus  $\vert \Delta \vert$ and
the critical temperature $T_{BKT}$. These two equations arise from 
the saddle-point and vortex actions.

To relate $\mu$ and $n = N/L^2$, where $N$ is the total number of particles per band
and $L^2$ is the area of the sample, we use
$
n = -\partial
{\overline  \Omega}/ \partial\mu \vert_{T, V},
$
with ${\overline \Omega} = \Omega/L^2$, 
leading to 
\begin{equation}
\label{eqn:number}
n =  n_{\rm sp} + n_{\rm ph} + n_{\rm zp}.
\end{equation}
Here, 
$
n_{j} = - \partial {\overline \Omega}_{j} / \partial \mu \vert_{T,V}$
where $j = \{{\rm sp}, {\rm ph}, {\rm zp} \}$. The expression of
$n_{\rm sp}$ is particularly simple, that is, 
$n_{\rm sp}= 2 \sum_{\bf k} n_{\rm sp} ({\bf k})$, while the others
can be easily obtained from their respective ${\overline \Omega}_j$.
The self-consistent solutions of Eqs.~(\ref{eqn:order-parameter}),
(\ref{eqn:critical-temperature}),
and (\ref{eqn:number}) determine $\mu$, $\vert \Delta \vert$ and $T_{\rm BKT}$ as
function of density $n$ for given effective mass $m$, interaction strength $V_{s}$ and
momentum-space interaction range $k_R$. A measure of the evolution from
BCS to Bose pairing is provided by the pair size~\cite{duncan-2000}
\begin{equation}
\label{eqn:pair-size}
\xi_{\rm pair}^2
=
\frac{\int d^2 {\bf k}\, \varphi^*({\bf k}) \left[ - \nabla^2_{\bf k} \right] \varphi ({\bf k})}
{ \int d^2 {\bf k}\, \vert \varphi ({\bf k}) \vert^2}
\end{equation}
in terms of the non-normalized two-body wavefunction
$\varphi ({\bf k}) = \Delta ({\bf k})/ 2 E ({\bf k})$.

We use momentum and energy scales $k_R$ and $\varepsilon_{R} = k_R^2/2m$,
respectively, and write the dimensionless order parameter equation as
\begin{equation}
\label{eqn:order-parameter-dimensionless}
\frac{1}{{\widetilde V}_s}
=
\int d{\tilde k}\, {\tilde k}
\frac{\vert {\widetilde \Gamma}_s ( {\tilde {\bf k}} ) \vert^2}{2 {\widetilde E}_{\tilde {\bf k}}}
\tanh \left( \frac{ {\widetilde E}_{\tilde {\bf k}}}{2 {\widetilde T}} \right),
\end{equation}
where ${\widetilde V}_s = V_s g_{2D}$, with $g_{2D} = m L^2/\pi$ being the total
2D density of states per band. The dimensionless functions
are ${\widetilde \Gamma}_s ( {\tilde {\bf k}}) = ( 1 + {\tilde k})^{-1/2}$ for the interaction
symmetry factor, 
$
{\widetilde E}_{\tilde {\bf k}}
=
\left[ {\tilde \xi}_{\tilde {\bf k}}^2 + \vert {\widetilde \Delta}_{\tilde {\bf k}}\vert^2 \right]^{1/2}
$
for the quasiparticle excitation energy, with ${\tilde \xi}_{\tilde {\bf k}} = {\tilde k}^2 - {\tilde \mu}$
for the kinetic energy and
$
{\widetilde \Delta}_{\tilde {\bf k}}
= {\widetilde \Delta} {\widetilde \Gamma}_s (\tilde {\bf k})
$
for the order parameter function.
The dimensionless critical temperature is
\begin{equation}
\label{eqn:critical-temperature-dimensionless}
{\widetilde T}_{\rm BKT}
=
\frac{{\tilde n}_{\rm sp}}{8}
-
\frac{1}{16 {\widetilde T}_{\rm BKT}}
\int d {\tilde k}\, {\tilde k}^3
{\rm sech}^2
\left(
\frac{{\widetilde E}_{\tilde {\bf k}}}{2 {\widetilde T}_{\rm BKT}}
\right),
\end{equation}
where ${\widetilde T}_{\rm BKT} = T_{\rm BKT}/T_R$, with $T_R = \varepsilon_R$.
The dimensionless density ${\tilde n} = n/n_R$, where $n_R = k_R^2/2\pi$,
is expressed solely in terms of the saddle-point contribution
\begin{equation}
\label{eqn:number-saddle-point-dimensionless}
\tilde n = {\tilde \varepsilon}_F = \tilde n_{\rm sp} =
\int d{\tilde k}\, {\tilde k}
\left[
1 - \frac{{\tilde \xi}_{\tilde {\bf k}}}{{\widetilde E}_{\tilde {\bf k}}}
\tanh \left( \frac{ {\widetilde E}_{\tilde {\bf k}}}{2 {\widetilde T}} \right)
\right],
\end{equation}
since $n_{\rm ph}$ and $n_{\rm zp}$
are small for $T \approx T_{\rm BKT}$,
except when $n$ and/or $T \to 0$~\cite{footnote-2}.
For comparison, we also obtain the dimensionless mean field (MF) temperature
${\widetilde T}_{\rm MF} = T_{\rm MF}/T_R$
by setting $\vert {\widetilde \Delta} \vert = 0$
and solving only Eqs.~(\ref{eqn:order-parameter-dimensionless})
and~(\ref{eqn:number-saddle-point-dimensionless}),
because MF neglects all phase fluctuations including the vortex-antivortex
binding mechanism described by Eq.~(\ref{eqn:critical-temperature-dimensionless}).

\begin{figure}[t]
\begin{center}
\includegraphics[width=0.98\linewidth]{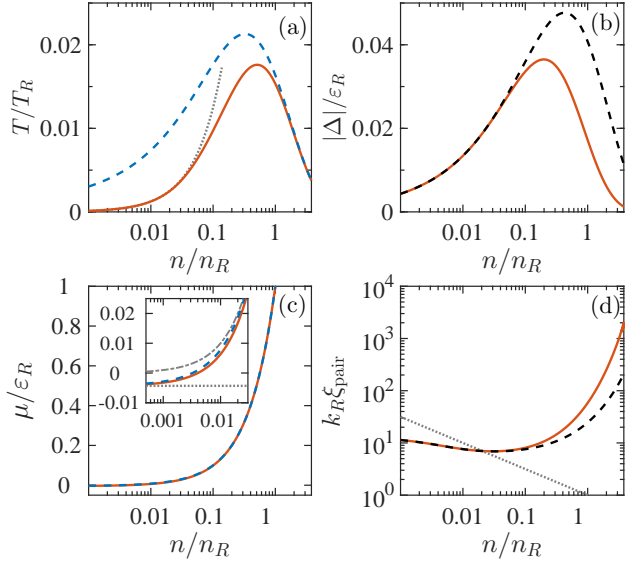}
\end{center}
\vspace{-8mm}
\caption{The coupling constant is ${\widetilde V}_s = 0.7100$ in all panels.
(a) Temperatures ${\widetilde T}_{{\rm BKT}}$ (red solid line),
${\widetilde T}_{\rm MF}$ (blue dashed line)  and upper bound
${\widetilde T}_{\rm BKT} = {\tilde \varepsilon}_{F}/8 = {\tilde n}/8$ (gray dotted line)
 vs ${\tilde n}$.
(b) Order parameter $\vert {\widetilde \Delta} \vert$
vs ${\tilde n}$ at ${\widetilde T} = {\widetilde T}_{\rm BKT}$ (red solid line)
and ${\widetilde T} = 0$ (black dashed line).
(c) Chemical potential ${\tilde \mu}$ vs ${\tilde n}$ at ${\widetilde T} = {\widetilde T}_{\rm BKT}$
(red solid line) and ${\widetilde T} = T_{\rm MF}$ (blue dashed line).
Inset: ${\tilde \mu} = {\tilde \varepsilon}_F = {\tilde n}$ (gray dot-dashed line)
and ${\tilde \mu} = - \vert {\widetilde E}_{B}\vert/2$ (gray dotted line).
(d) Pair size ${\tilde \xi}_{\rm pair}$ vs ${\tilde n}$ at
${\widetilde T} = {\widetilde T}_{\rm BKT}$ (red solid line),
${\widetilde T} = 0$ (black dashed line), and
${\tilde \xi}_{\rm pair} = {\tilde n}^{-1/2}$ (gray dotted line), that is,
$k_F \xi_{\rm pair} = 1$ at ${\widetilde T} = 0$.
}
\label{fig:one}
\end{figure}

In Fig.~\ref{fig:one}, we show temperatures ${\widetilde T}_{\rm BKT}= T_{\rm BKT}/T_R $,
${\widetilde T}_{\rm MF} = T_{\rm MF}/T_R$,
order parameter $\vert {\widetilde \Delta}\vert = \vert \Delta \vert / \varepsilon_R$,
chemical potential $\tilde \mu = \mu / \varepsilon_R$ and
pair size ${\tilde \xi}_{\rm pair} =  k_R \xi_{\rm pair}$ versus
density ${\tilde n} = n/ n_R$ for fixed ${\widetilde V}_s = 0.7100$.
Notice that 
${\tilde n} = {2\pi} (R/\ell_{\rm ip})^2$
and a crossover
from short- to long-ranged interactions occurs at $R/\ell_{\rm ip} \sim 1$, that is, at
${\tilde n} \sim {\tilde n}_{\rm cr} = 2\pi = 6.283$ as ${\tilde n}$ grows. 
In Fig.~\ref{fig:one}(a), we show ${\widetilde T}_{\rm BKT}$ and ${\widetilde T}_{\rm MF}$
versus ${\tilde n}$ and note that
${\widetilde T}_{\rm MF} \ge {\widetilde T}_{\rm BKT}$.
The gray dotted line represents the upper bound
${\widetilde T_{\rm BKT}} = {\tilde n}/8 = {\tilde \varepsilon}_F/8$ discussed
in the introduction~\cite{botelho-2006}.
Both ${\widetilde T}_{\rm MF}$ and ${\widetilde T}_{\rm BKT}$ are non-monotonic and
have a maximum located at ${\tilde n} = 0.3241$ and ${\tilde n} = 0.5099$, respectively.
The inclusion of phase fluctuations, via the BKT mechanism of vortex-antivortex
binding, reduces $T_{\rm MF}$ to $T_{\rm BKT}$ through
Eq.~(\ref{eqn:critical-temperature-dimensionless}).
Therefore the range  $T_{\rm BKT} < T < T_{\rm MF}$ determines
the region where classical
phase fluctuations are important. In this region, preformed pairs are
guaranteed to exist~\cite{sademelo-1993, sademelo-2008}. However, we cannot interpret
$T_{\rm MF}$ as the experimental pseudogap temperature $T^*$ defined
as a $1\%$ drop in the zero-bias conductance~\cite{iwasa-2021}.
The theoretical understanding of $T^*$ requires further analysis~\cite{footnote-3}.

In Fig.~\ref{fig:one}(b), we show the $\vert {\widetilde \Delta} \vert$ versus ${\tilde n}$
both at ${\widetilde T} = {\widetilde T}_{\rm BKT}$ and ${\widetilde T} = 0$,
noticing that $\vert {\widetilde \Delta}\vert$ at ${\widetilde T} = {\widetilde T}_{\rm BKT}$
has a maximum at ${\tilde n} = 0.1939$, while $\vert {\widetilde \Delta}  \vert$
at ${\widetilde T} = 0$ has a maximum at ${\tilde n} = 0.4356$.
In Fig.~\ref{fig:one}(c), we show ${\tilde \mu}$ versus ${\tilde n}$
both at ${\widetilde T} = {\widetilde T}_{\rm BKT}$ and ${\widetilde T} = 0$,
observing that ${\tilde \mu}$ is a monotonically increasing function of
${\tilde n}$ as expected.
For higher densities, where the superconductor is deep in the degenerate
BCS regime, $\tilde \mu \to {\tilde \varepsilon}_F = {\tilde n}$,
as indicated by the gray dot-dashed line.
For lower densities, where the superconductor
is deep in the non-degenerate Bose regime,
${\tilde \mu} \to  - \vert {\widetilde E}_B \vert/2 = -0.004309$ 
as indicated by the gray dotted line.
Here, ${\widetilde E}_{\rm B} = - 0.008619$ is the dimensionless binding energy
obtained from the two-body bound state eigenvalue
equation~\cite{botelho-2006}
\begin{equation}
\frac{1}{{\widetilde V}_s}
=
\int d{\tilde k}\, {\tilde k}
\frac{\vert {\widetilde \Gamma}_s ( {\tilde {\bf k}} ) \vert^2}{2 {\tilde k}^2 - {\widetilde E}_B}.
\end{equation}
for ${\widetilde V}_s = 0.7100$.
Spectroscopically, the BCS-Bose crossover occurs
at ${\tilde \mu} = 0$ $({\tilde n} = 0.004032)$,
where the gap for quasiparticle excitations 
changes from ${\widetilde E}_{\rm g}= \vert {\widetilde \Delta} \vert$
on the BCS-pairing side $({\tilde n} > 0.004032)$ to 
${\widetilde E}_{\rm g} = \sqrt{{\tilde \mu}^2 + \vert {\widetilde \Delta} \vert^2}$
on the Bose-pairing side $({\tilde n} < 0.004032)$.

In Fig.~\ref{fig:one}(d), we show that the pair size
${\tilde \xi}_{\rm pair} = k_R \xi_{\rm pair}$ is not a monotonically increasing function
of ${\tilde n}$, having a minimum at ${\tilde n} = 0.03275$.
We emphasize that $\xi_{\rm pair}$ is not the phase coherence length
$\xi_{\rm ph}$ that diverges at the ${\rm BKT}$ transition, but it provides a measure 
that the BCS-Bose pairing crossover region occurs when $k_F \xi_{\rm pair} \sim 1$~\cite{sademelo-1993}.
The condition $k_F \xi_{\rm pair} = 1$
or ${\tilde \xi}_{\rm pair} = {\tilde k}_F^{-1} = {\tilde n}^{-1/2}$
is represented by the gray dotted line.  
The BCS-pairing limit corresponds to 
${\tilde \xi}_{\rm pair} \gg {\tilde n}^{-1/2}$, that is, $k_F \xi_{\rm pair} \gg 1$.
In contrast, the Bose-pairing limit corresponds
to ${\tilde \xi}_{\rm pair} \ll {\tilde n}^{-1/2}$, that is,  $k_F \xi_{\rm pair} \ll 1$.

\begin{figure}[t]
\begin{center}
\includegraphics[width=0.98\linewidth]{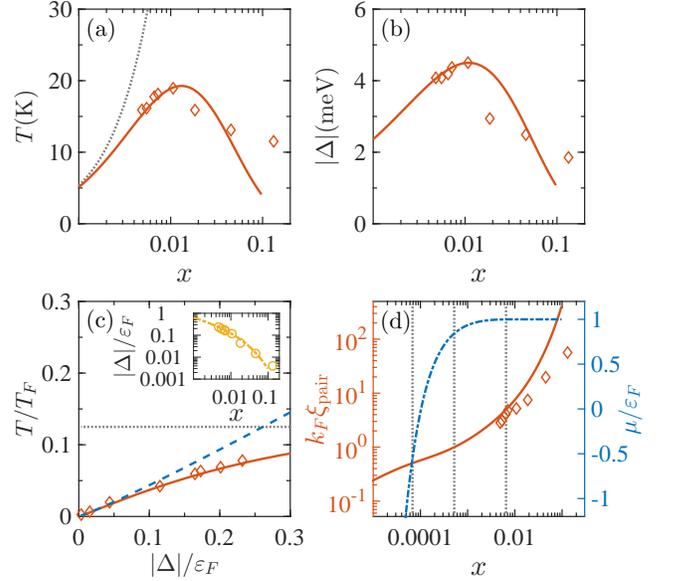}   
\end{center}
\vspace{-8mm}
\caption{Theoretical parameters are
 ${\widetilde V}_s = 0.7100$, $T_R = 1094 {\rm K}$, $k_R = 0.4310/a$.
Experimental data~\cite{iwasa-2021} (red diamonds and yellow circles).
(a) $T$ (in K) vs $x$, theoretical $T_{\rm BKT}$ (red solid line). 
(b) $\vert \Delta \vert$ (in meV) at $T = 2{\rm K}$,
theoretical $\vert \Delta \vert$ (red solid line).
(c) $T/T_F$ vs $\vert \Delta \vert/\varepsilon_F$, 
$T_{\rm BKT}/T_F$ (red solid line),
$T_{\rm MF}/T_F$ (blue dashed line), and
$T_{\rm BKT}/T_F = 1/8 = 0.125$ (gray dotted line).
Inset: $\vert \Delta \vert$ vs $x$, theoretical (yellow dot-dashed line).
(d) Theoretical $k_F \xi_{\rm pair}$ (red solid line), experimental $k_F \xi_{\rm exp}$ (red diamonds) 
both at $T = 0$;  $\mu/\varepsilon_F$ (blue dot-dashed line) at $T = T_{\rm BKT}$.  
The vertical gray dotted lines at $x = \{6.549 \times 10^{-5}, 5.238 \times 10^{-4}, 6.564 \times 10^{-3} \}$
give $k_F \xi_{\rm pair} = \{0.5, 1.0, 5.0\}$. 
}
\label{fig:two}
\end{figure}

In Fig.~\ref{fig:two}, we compare our theoretical results for $T_{\rm BKT}$,
$\vert \Delta \vert$, $\mu$, and $\xi_{\rm pair}$
to experimental results for ${\rm Li}_{x}{\rm ZrNCl}$~\cite{iwasa-2021}.
In Figs.~\ref{fig:two}(a) and (b), we vary the dimensionless coupling
constant ${\widetilde V_s}$ to match the theoretical $T_{\rm BKT}$ and $\vert \Delta \vert$
to the experimental $T_c$ and $\vert \Delta \vert$ versus concentration $x$. The best fit is achieved for
${\widetilde V}_s = 0.7100$, and from it we extract
the temperature scale $T_R = \varepsilon_{R} = k_R^2/2m$ leading to $T_R = 1094{\rm K}$ 
and the momentum scale $k_R$ by relating ${\tilde n}$ to $x$, that is, by using
the carrier density per band $n = x/A_{\rm cell}$,
where $A_{\rm cell} = a^2 \sin(\pi/3)$ is the area of the unit cell of ${\rm Li}_{x}{\rm ZrNCl}$.
In this case, ${\tilde n} = 2\pi x/k_R^2 A_{\rm cell}$,
and the momentum scale from the fit is $k_R = 0.4310/a = 1.197{\rm nm}^{-1}$,
where $a = 0.3601{\rm nm}$ is the characteristic length of the unit cell.
Note that the interaction range
$R \sim 1/k_R = 2.320 a = 0.8354{\rm nm}$ spans a few lattice spacings,
suggesting that the extension to a lattice model should require interactions
involving only a few neighbors.
The effective mass $m$ is obtained from $m = k_R^2/2\varepsilon_R$ leading to
$m = 0.5791 m_e$, where $m_e$ is the bare electron mass, in good agreement
with band structure calculations~\cite{pickett-1999, heid-2005}.
Since $\tilde n = 2\pi (R/\ell_{\rm ip})^2$, the crossover from
short- to long-ranged interactions occurs at $R/\ell_{\rm ip} \sim 1$, that is, $\tilde n \sim {\tilde n}_{\rm cr} = 2\pi$ 
$(x \sim 0.1608)$, meaning that interactions are sufficiently short-ranged over the experimental range,
however a non-zero $R$ is absolutely essential in producing a maximum in
$T_{\rm BKT}$ versus $x$.

In Fig.~\ref{fig:two}(b), we show $\vert \Delta \vert$  (in ${\rm meV}$) at $T = 2{\rm K}$ versus $x$.
The red diamonds are the experimental data, and the red solid  line are theoretical
results using the same values of $T_R$, $k_R$ and ${\widetilde V}_s$ given in Fig.~\ref{fig:two}(a).
In Fig.~\ref{fig:two}(c), we show $T_{\rm BKT}/T_F$ and $T_{\rm MF}/T_F$ versus
$\vert \Delta \vert/\varepsilon_F$ at $T = 2{\rm K}$ for the same
values of $T_R$, $k_R$ and ${\widetilde V}_s$. The upper bound
$T_{\rm BKT} = 0.125 T_F$~\cite{botelho-2006}
is the gray dotted line. The range $T_{\rm BKT}/T_F < T/T_F < T_{\rm MF}/T_F$ determines the region
where phase fluctuations are important. The inset shows $\vert \Delta \vert /\varepsilon_F$ versus
$x$ comparing experiment (yellow circles) and theory (yellow dot-dashed line).
In Fig.~\ref{fig:two}(d), we show $k_F \xi_{\rm pair}$ and $\mu/\varepsilon_F$ versus $x$.
The theoretical $k_F \xi_{\rm pair}$ is compared with the
experimental $k_F \xi_{\rm exp}$~\cite{footnote-4} at $ T = 0$,
while the theoretical $\mu/\varepsilon_F$ is at $T = T_{\rm BKT}$. 
The vertical gray dotted lines at $x = \{6.549 \times 10^{-5}, 5.238 \times 10^{-4}, 6.564 \times 10^{-3} \}$
give $k_F \xi_{\rm pair} = \{0.5, 1.0, 5.0\}$.  The crossover region from BCS to Bose pairing occurs in
the range $0.5 < k_F \xi_{\rm pair} < 5.0$ or $6.549 \times 10^{-5} < x < 6.564 \times 10^{-3}$,
with corresponding $ -0.6113 < \mu/\varepsilon_F < 0.9995 $. At the lowest
experimental concentration $x = 0.0048$, we find $k_F \xi_{\rm pair} = 3.826 $ and
$\mu/\varepsilon_F = 0.9979$, showing that the crossover region has just been entered.
Our results demonstrate that the Bose pairing regime has not yet been reached experimentally
for gated 2D  ${\rm Li}_x {\rm ZrNCl}$, as it 
requires $x <  6.549 \times 10^{-5}$, $k_F\xi_{\rm pair} < 0.5$ and $\mu/\varepsilon_F < -0.6113$. 

{\it Final Remarks:} Our theoretical work shows good agreement with experimental
data in the range $0.0048 \le x \le 0.133$~\cite{iwasa-2021}. However, for $ x \ge 0.1$
there are some deviations which our current model does not capture as described next.
First, according to Ref.~\cite{iwasa-2021}, for $x \ge 0.1$ there is a crossover in behavior from
two to three dimensions, which our strictly 2D model does not include. Second,
it was suggested~\cite{iwasa-2021} that electron-phonon interactions may play a role specially
for $x \ge 0.1$, but our model does not include retardation effects. Third, for $x \ge 0.1$, our
results suggest that it is necessary to include band-structure effects (non-parabolicity of bands)
with a corresponding increase in the density of states that leads to an increase
in $T_c$, $\vert \Delta \vert$ and a decrease in $k_F \xi_{\rm pair}$.

{\it Conclusions:} We investigated a density induced evolution from BCS to Bose pairing
in gated two-dimensional superconductors with two degenerate parabolic bands and
showed that there is good agreement with experimental
results~\cite{iwasa-2021} of ${\rm Li}_{x}{\rm ZrNCl}$ for $x \le 0.1$, where the
effective mass approximation is valid.
We conclude that, at the lowest reported concentration
$x = 0.0048$, the chemical potential is still very close to the Fermi energy, and that the
pair size is $k_F \xi_{\rm pair} = 3.826$, indicating that experiments have just entered the
crossover region from the BCS-side. We suggest that it is necessary to reduce the
lowest concentration by at least one order of magnitude to start entering the
Bose pairing regime experimentally.

\acknowledgments{We thank Yoshiro Iwasa for discussions, and 
the National Key R$\&$D Program of China (Grant 2018YFA0306501),
the National Natural Science Foundation of China (Grants 11522436 $\&$ 11774425),
the Beijing Natural Science Foundation (Grant Z180013), and the Research Funds of
Renmin University of China (Grants 16XNLQ03 $\&$18XNLQ15) for financial support.}



\begin{thebibliography}{2}

\bibitem{goldman-2013}
A. M. Goldman, 
The Berezinskii-Kosterlitz-Thouless transition in superconducors,
In 40 years of the Berezinskii-Kosterlitz-Thouless theory,
135, World Scientific (2013).
  
\bibitem{berezinskii-1970}
V. L. Berezinskii,
Destruction of long-range order in one-dimensional and two-dimensional systems having
a continuous symmetry group : I. Classical systems.   
Sov. Phys. JETP {\bf 32}, 493 (1970).

\bibitem{kosterlitz-thouless-1972}
J. M. Kosterlitz and D. Thouless,
Long-range order and metastability in two dimensional solids and superfluids: Application of
dislocation theory.
J. Phys. C {\bf 5}, L124 (1972)

\bibitem{beasley-1979}
M. R. Beasley, J. E. Mooij, and T. P. Orlando, Phys
Possibility of Vortex-antivortex pair dissociation in two-dimensional superconductors,
Rev. Lett. {\bf 42}, 1165 (1979).

\bibitem{goldman-1981}
K. Epstein, A. M. Goldman, and A. M. Kadin,
Vortex-antivortex pair dissociation in two-dimensional superconductors,
Phys. Rev. Lett. {\bf 47}, 534 (1981).

\bibitem{halperin-1979}
B. I. Halperin and D. R. Nelson,
Resisitive transition in superconducting films,
J. Low Temp. Phys. {\bf 36}, 599 (1979).

\bibitem{duncan-2000}
R. D. Duncan and C. A. R. S{\' a} de Melo, 
Thermodynamic properties in the evolution from BCS to Bose-Einstein condensation
for a d-wave superconductor at low temperatures,
Phys. Rev. B {\bf 62}, 9675 (2000).

\bibitem{botelho-2005a}
S. S. Botelho and C. A. R. S{\'a} de Melo,
Lifshitz transition in d-wave superconductors,
Phys. Rev B {\bf 71}, 134507 (2005).

\bibitem{read-2000}
N. Read and D. Green,
Paired states of fermions in two dimensions with breaking of parity
and time-reversal symmetries and the fractional quantum Hall effect,
Phys. Rev. B {\bf 61}, 10267 (2000)
  
\bibitem{botelho-2005b}
S. S. Botelho and C. A. R. S\'{a} de Melo,
Quantum Phase Transition in the BCS-to-BEC Evolution of p-wave Fermi Gases,
J.  Low Temp.  Phys. {\bf 140}, 409 (2005).

\bibitem{ahn-2003a}
C. H. Ahn, J.-M. Triscone, and J. Mannhart,
Electric field effect in correlated oxide systems,
Nature {\bf 424}, 1015, (2003).

\bibitem{ahn-2003b}
D. Matthey, S. Gariglio, C.H. Ahn, and J.-M. Triscone,
Electrostatic modulation of the superconducting transition in thin
${\rm NdBa_2Cu_3O_{7-\delta}}$ films:
The role of classical fluctuations.
Physica C {\bf 583}, 372, (2002).

\bibitem{kivelson-1995}
V. J. Emery and S. A. Kivelson, 
Importance of phase fluctuations in superconductors
with small superfluid density,
Nature {\bf 374}, 434 (1995).

\bibitem{uemura-1989}
Y. J. Uemura, G. M. Luke,  B. J. Sternlieb, J. H. Brewer, J. F. Carolan, W. N. Hardy,
R. Kadono, J. R. Kempton, R. F. Kiefl, S. R. Kreitzman, P. Mulhern, T. M. Riseman,
D. Ll. Williams, B. X. Yang, S. Uchida, H. Takagi, J. Gopalakrishnan, A. W. Sleight,
M. A. Subramanian, C. L. Chien, M. Z. Cieplak, Gang Xiao, V. Y. Lee, B. W. Statt,
C. E. Stronach, W. J. Kossler, and X. H. Yu,
Universal Correlations between T, and $n_s /m^*$ (carrier density over effective mass)
in high-$T_c$ cuprate superconductors,
Phys. Rev. Lett. {\bf 62},  2317  (1989).

\bibitem{uemura-1991}  
Y. J. Uemura,  L. P. Le,  G. M. Luke,  B. J. Sternlieb,  W. D. Wu,  J. H. Brewer, T. M.
Riseman, C. L. Seaman, M. B. Maple, M. Ishikawa, D. G. Hinks, J. D. Jorgensen, G.
Saito, and H. Yamochi,
Basic similarities among cuprate, bismuthate, organic, Chevrel-phase, and heavy-fermion
superconductors shown by penetration-depth measurements,
Phys. Rev. Lett. {\bf 66},  2665 (1991).

\bibitem{botelho-2006}
S. S. Botelho and C. A. R. S\'{a} de Melo,
Vortex-Antivortex Lattice in Ultracold Fermionic Gases,
Phys. Rev. Lett.  {\bf 96}, 040404 (2006).

\bibitem{ferrell-1958}
R. A. Ferrell and R. E. Glover,
Conductivity of superconducting films: A sum rule,
Phys. Rev. {\bf 109}, 1398 (1958).
  
\bibitem{tinkham-1959}
M. Tinkham and R. A. Ferrell,
Determination of the superconducting skin depth from the energy gap and sum rule,
Phys. Rev. Lett. {\bf 2}, 331 (1959).

\bibitem{tinkham-1975}
M. Tinkham,
Introduction to Superconductivity,
McGraw-Hill, New York, 1975.

\bibitem{devreese-2014}
J. P. A. Devreese, J. Tempere and Carlos A. R. S{\'a} de Melo,
Effects of Spin-Orbit Coupling on the Berezinskii-Kosterlitz-Thouless Transition
and the Vortex-Antivortex Structure in Two-Dimensional Fermi Gases,
Phys. Rev. Lett. {\bf 113}, 165304 (2014).

\bibitem{devreese-2015}
J. P. A. Devreese, J. Tempere, and C. A. R. S{\'a} de Melo,
Quantum phase transitions and Berezinskii-Kosterlitz-Thouless temperature
in a two-dimensional spin-orbit-coupled Fermi gas,
Phys. Rev. A {\bf 92}, 043618 (2015).

\bibitem{kasahara-2014}
S. Kasahara, T. Watashige, T. Hanaguri, Y. Kohsaka, T. Yamashita, Y. Shimoyama, Y. Mizukami, R. Endo,
H. Ikeda, K. Aoyama, T. Terashima, S. Uji, T. Wolf, H. von Löhneysen, T. Shibauchi, and Y. Matsuda,
Field-induced superconducting phase of FeSe in the BCS-BEC crossover,
Proc. Natl. Acad. Sci. U.S.A. {\bf 111}, 16309–16313 (2014).
   
\bibitem{rinott-2017}
S. Rinott, K. B. Chashka, A. Ribak, E. D. L. Rienks, A. Taleb-Ibrahimi, P. Le Fevre, F. Bertran, M. Randeria
and A. Kanigel, Tuning across the BCS-BEC crossover in the multiband superconductor
${\rm Fe}_{1+y} {\rm Se}_x  {\rm Te}_{1-x}$: An angle-resolved photoemission study,
Sci. Adv. {\bf 3}, e1602372 (2017).

\bibitem{hashimoto-2020}
T. Hashimoto, Y. Ota, A. Tsuzuki, T. Nagashima, A. Fukushima, S. Kasahara, Y. Matsuda, K. Matsuura,
Y. Mizukami, T. Shibauchi, S. Shin, and K. Okazaki,
Bose-Einstein condensation superconductivity induced by disappearance of the nematic state,
Sci. Adv. {\bf 6}, eabb9052 (2020).

\bibitem{cao-2018}
Y. Cao, V. Fatemi, S. Fang, K. Watanabe, T. Taniguchi, E. Kaxiras, and P. Jarillo-Herrero,
Unconventional superconductivity in magic-angle graphene superlattices,
Nature {\bf 556}, 43 (2018).
  
\bibitem{herrero-2021}
J. M. Park, Y. Cao, K. Watanabe, T. Taniguchi, P. Jarillo-Herrero,
Tunable strongly coupled superconductivity in magic-angle twisted trilayer graphene,
Nature {\bf 590}, 249–255 (2021).

\bibitem{nakagawa-2018}
Y. Nakagawa, Y. Saito, T. Nojima, K. Inumaru, S. Yamanaka, Y. Kasahara, and Y. Iwasa,
Gate-controlled low carrier density superconductors: Toward the two-dimensional BCS-BEC crossover
Phys. Rev. B {\bf 98}, 064512 (2018).

\bibitem{iwasa-2021}
Yuji Nakagawa, Yuichi Kasahara, Takuya Nomoto, Ryotaro Arita, Tsutomu Nojima,
Yoshihiro Iwasa, Gate-controlled BCS-BEC crossover in a two-dimensional superconductor,
Science {\bf 372}, 190–195 (2021).
  
\bibitem{footnote-1} This parametrization is necessary to produce a
separable potential in momentum space that behaves both at small and large momenta
in a form that is compatible to an acceptable real space interaction
potential $V({\bf r}, {\bf r}^{\prime})$, as previously investigated
in the literature~\cite{duncan-2000}.

\bibitem{footnote-2} Additional contributions to the number equation
beyond phase fluctuations also arise
in the Bose limit (low densities)
and lead to logarithmic corrections of $T_{\rm BKT}$
due to residual boson-boson interactions. See, e.g.,
D. S. Fisher, and P. C. Hohenberg,
Dilute Bose gas in two dimensions,
Phys. Rev. B {\bf 37}, 4936 (1988).

\bibitem{sademelo-1993}
C. A. R. S{\' a} de Melo, M. Randeria, and J. R. Engelbrecht, 
Crossover from BCS to Bose superconductivity: Transition temperature and time-dependent Ginzburg-Landau theory,
Phys. Rev. Lett. {\bf 71}, 3202 (1993).

\bibitem{sademelo-2008}
 C. A. R. S\'{a} de Melo,
 When fermions become bosons: Pairing in ultracold gases,
 Physics Today, October issue, 45 (2008).

\bibitem{footnote-3}
For $T > T_{\rm MF}$, where $\vert \Delta \vert = 0$, fermions still interact and cannot be
treated as free particles. Since there is always a two-body bound state in two dimensions
for an arbitrary small attractive interaction, these bound states must 
be included for $T > T_{\rm MF}$ and the density of fermions needs to separated at least
into bound (preformed pairs) and unbound contributions.
To recover bound fermions above $T_{\rm MF}$ it is necessary to include
amplitude fluctuations of $\Delta$. Therefore, $T^*$ is related to the
Saha dissociation temperature $T_{\rm S}$, where a fraction of preformed pairs $B$
dissociates into two fermions $F_\uparrow$ and $F_\downarrow$ with opposite spins,
satisfying the chemical equilibrium equation $B \rightleftharpoons \, F_\uparrow + F_\downarrow$.
Details of this analysis will be investigated in a future publication.
  
\bibitem{pickett-1999}
R. Weht, A. Filippetti, W. E. Pickett,
Electron doping in the honeycomb bilayer superconductors ${\rm (Zr, Hf) NCl}$,
Europhys. Lett. {\bf 48}, 320–325 (1999).

\bibitem{heid-2005}
R. Heid, K. P. Bohnen,
Ab Initio lattice dynamics and electron-phonon coupling in ${\rm Li}_x {\rm Zr N Cl}$.
Phys. Rev. B {\bf 72}, 134527 (2005).

\bibitem{footnote-4}
The experimental parameter
$\xi_{\rm exp}$ is extracted from the measurement of the temperature dependent of the
upper critical field $H_{c_2}(T)$ using the Ginzburg-Landau relation
$H_{c_2} (T) = \Phi_0/2\pi\xi^2(T)$, with $\xi (T) = \xi_{\rm exp} (1 - T/T_c)^{-1/2}$ and
extrapolating it to $T = 0$,
from $H_{c_2} (T) = \left( \Phi_0/2\pi\xi^2_{\rm exp}\right) (1 - T/T_c)$ for $T < T_c$.
Here $\Phi_0$ is the flux quantum $\hbar c/2e$,
and $T_c$ is the zero-field critical temperature. The experimental value
$\xi_{\rm exp} = \left[ 2\pi H_{c_2} (0)/\Phi_0\right]^{1/2}$ and the theoretical values of
$\xi_{\rm pair}$ are in good agreement in the range $0 \le x \le 0.1$, where
the system is expected to be in the truly 2D regime~\cite{iwasa-2021},
band structure (lattice) effects are not important, and the experimental system is closer
to the BCS-pairing side of the crossover. At the Bose-pairing side of the crossover is approached,
we expect substantial deviations between $\xi_{\rm pair}$ and $\xi_{\rm exp}$ as discussed in the
three-dimensional version of the BCS-BEC crossover~\cite{sademelo-1993}.
\end{thebibliography}
\end{document}